\documentstyle[aps,multicol,tabularx,graphics,rotate,epsf]{revtex}
\begin{document}
\title{Roughening Transition in a Moving Contact Line}
\author{Ramin Golestanian$^{1,2,3}$ and Elie Rapha\"el$^{1}$}
\address{$^{1}$ Laboratoire de Physique de la Mati\`ere Condens\'ee,
College de France, UMR 7125 et FR 2438 du CNRS, 11 place
Marcelin-Berthelot,
75231 Paris Cedex 05, France\\
$^{2}$Institute for Advanced Studies in Basic Sciences, Zanjan
45195-159, Iran \\
$^{3}$Institute for Studies in Theoretical Physics and
Mathematics, P.O. Box 19395-5531, Tehran, Iran}
\date{\today}
\maketitle
\begin{abstract}
The dynamics of the deformations of a moving contact line on a
disordered substrate is formulated, taking into account both local
and hydrodynamic dissipation mechanisms. It is shown that both the
coating transition in contact lines receding at relatively high
velocities, and the pinning transition for slowly moving contact
lines, can be understood in a unified framework as roughening
transitions in the contact line. We propose a phase diagram for
the system in which the phase boundaries corresponding to the
coating transition and the pinning transition meet at a {\it
junction} point, and suggest that for sufficiently strong disorder
a receding contact line will leave a Landau--Levich film
immediately after depinning. This effect may be relevant to a
recent experimental observation in a liquid Helium contact line on
a Cesium substrate [C. Guthmann, R. Gombrowicz, V. Repain, and E.
Rolley, Phys. Rev. Lett. {\bf 80}, 2865 (1998)].

\end{abstract}

\begin{multicols}{2}

\section{Introduction an Summary}  \label{sIntro}

When a drop of liquid spreads on a solid surface, the {\it contact
line}, which is the common borderline between the solid, the
liquid, and the corresponding equilibrium vapor, undergoes a
rather complex dynamical behavior. This dynamics is determined by
a subtle competition between the mutual interfacial energetics of
the three phases, dissipation and hydrodynamic flows in the
liquid, and the geometrical or chemical irregularities of the
solid surface \cite{dG1}.

For a partially wetting fluid on sufficiently smooth substrates, a
contact line at equilibrium has a well defined contact-angle
$\theta_e$ that is determined by the solid-vapor $\gamma_{\rm SV}$
and the solid-liquid $\gamma_{\rm SL}$ interfacial energies, and
the liquid surface tension $\gamma$ through Young's relation:
$\gamma_{\rm SV}-\gamma_{\rm SL}=\gamma \cos \theta_e$.
For a moving contact line, however, the value of the so-called
dynamic contact-angle $\theta$ changes as a function of velocity:
$\theta > \theta_e$ for an advancing contact line and $\theta <
\theta_e$ for a receding one. This is because the unbalanced
interfacial force $\gamma_{\rm SV}-\gamma_{\rm SL}-\gamma \cos
\theta$ now has to be balanced with a frictional force in a steady
state situation. The dissipation in the moving contact line, which
is responsible for the friction, can be either of {\em local}
origin, for example due to microscopic jumps of single molecules
(from the liquid into the vapor) in the immediate vicinity of the
contact line \cite{Blake,BR}, or due to viscous {\em hydrodynamic}
losses inside the moving liquid wedge
\cite{dG1,Voinov,Cox,hydro,hydro1}.

For a contact line that is receding at a velocity $v$, it has been
shown by de Gennes \cite{hydro} that a steady state is achieved in
which the liquid will partially wet the plate with a nonvanishing
dynamic contact-angle $\theta$ only for velocities less than a
certain critical value. The dynamic contact-angle decreases with
increasing $v$, until at the critical velocity the system
undergoes a dynamical phase transition in which a macroscopic
Landau--Levich liquid film \cite{LL,Levich}, formally
corresponding to a vanishing $\theta$, will remain on the plate.
One can think of the dynamic contact-angle as the order parameter
characterizing this {\em coating transition}, in analogy with
equilibrium phase transitions, while velocity is playing the role
of the tuning parameter. Elaborating further on this analogy then
seems to suggest that the nature of the coating transition depends
crucially on the dominant dissipation mechanism: In the local
picture $\theta$ vanishes continuously as $v$ approaches the
critical velocity, which makes it look like a {\em second order}
phase transition, while on the contrary, in the hydrodynamic
picture a jump is predicted in $\theta$ from $\theta_e/\sqrt{3}$
to zero at the transition, which is the signature of a {\em first
order} phase transition \cite{hydro}.

Another notable feature of contact lines, which is responsible for
their novel dynamics, is their anomalous long-ranged elasticity
\cite{JdG1}. For length scales below the capillary length (which
is of the order of 3 mm for water at room temperature), a contact
line deformation of wavevector $k$ will distort the surface of the
liquid over a distance $|k|^{-1}$. Assuming that the surface
deforms instantaneously in response to the contact line, the
elastic energy cost for the deformation can be calculated from the
surface tension energy stored in the distorted area, and is thus
proportional to $|k|$. The anomalous elasticity leads to
interesting equilibrium dynamics, corresponding to when the
contact line is perturbed from its static position, as studied by
de Gennes \cite{dG2}. Balancing the rate of interfacial energy
change and the dissipation, which he assumed for small
contact-angles is dominated by the hydrodynamic dissipation in the
liquid nearby the contact line, he finds that each deformation
mode relaxes to equilibrium with a characteristic (inverse) decay
time $\tau^{-1}(k)=c |k|$, where $c$ is a characteristic {\em
relaxation velocity} \cite{dG2}. The relaxation is thus
characterized by a dynamic exponent $z$, defined via $\tau^{-1}(k)
\sim |k|^z$, which is equal to 1. The linear dispersion relation
implies that a deformation in the contact line will {\it decay}
and {\it propagate} at a constant velocity, as opposed to systems
with normal line tension elasticity, where the decay and the
propagation are governed by diffusion. This behavior has been
observed, and the linear dispersion relation has been precisely
tested, in a recent experiment by Ondarcuhu and Veyssie
\cite{exp1}.

In reality, the presence of defects and heterogeneities in the
substrate, which could be due to (surface) roughness or chemical
contamination, further complicates the dynamics of a contact line
\cite{diMeglio,Fermigier}. In the presence of such
heterogeneities, a contact line at equilibrium becomes {\it
rough}, because it tries to locally deform so as to find the path
with optimal {\it pinning} energy \cite{dG1}. This is in contrast
to the case of a perfect solid surface, where the contact line is
{\it flat}. The roughness can be characterized as a scaling law
that relates the statistical width $W$ of the contact line to its
length $L$, via $W \sim L^{\zeta}$. The so-called roughness
exponent $\zeta$ is equal to $1/3$ for a contact line at
equilibrium on a surface with short-range correlated disorder
\cite{Huse,RJ1,mezard,Rolley}. Since the contact line is pinned by
the defects, a nonzero (critical) force is necessary to set it to
motion, through a {\em depinning} transition
\cite{Elie,JoaRob,Ertas,Wong}. For a contact line at the depinning
threshold, a roughness exponent about $0.4$ has been predicted
theoretically from two-loop field theoretical renormalization
group calculations \cite{LeDoussal}, and numerical simulations
\cite{Krauth}, which seems to disagree with the experimental
finding of $0.5$ for both liquid Helium on a Cesium substrate
\cite{Rolley2} and water on glass experiments \cite{Rolley3}. It
is also important to note that there may be numerous metastable
states for the contact line due to the random disorder, leading to
hysteresis in the contact-angle \cite{JdG1,RJ1}.

\vskip-2.5truecm
\begin{figure}
\centerline{\epsfxsize 7.5cm {\epsffile{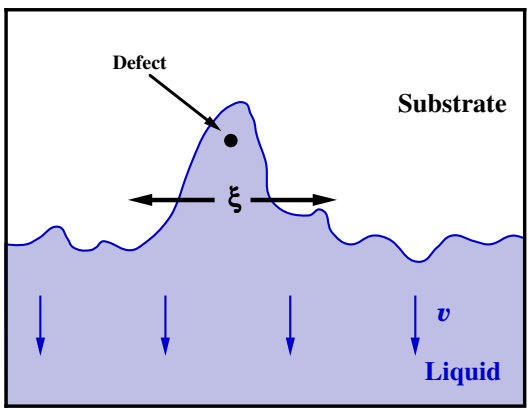}}}
\vskip.3truecm \caption{A contact line moving on a disordered
substrate undergoes shape fluctuations. A portion of the contact
line can be instantaneously pinned by a defect, thereby nucleating
domains of typical sizes given by a correlation length $\xi$.
These domains become rough, because they have to conform to the
minimum energy configuration on the substrate. At the onset of a
roughening transition this correlation length diverges.}
\label{fig:xi}
\end{figure}

Here we study the {\em nonequilibrium} dynamics of the
deformations of a moving contact line on a disordered substrate
\cite{RE-EPL,RE-PRE}. The dynamics is governed by a balance
between three different forces: (i) the interfacial force, (ii)
the frictional force which can stem from either local or
hydrodynamic dissipation processes, and (iii) a random force
caused by the disorder. We find that the relaxation spectrum of a
moving contact line is the same as the equilibrium case, but the
characteristic relaxation velocity depends on $v$: It decreases
with $v$ until at the critical velocity corresponding to the
coating transition it vanishes identically. The progressively slow
relaxation of a distorted contact line near the coating transition
is in agreement with a nucleation picture of the phase transition
(See Fig.~\ref{fig:xi}).

We find that coating transition can be actually understood in
terms of a {\em roughening transition} of the contact line on the
disordered substrate. Since linear relaxation becomes infinitely
slow in the vicinity of the coating transition, the dominant
relaxation is thus governed by nonlinear terms in the dynamical
equation, and the dynamical phase transition can thus be properly
accounted for only by using systematic renormalization group (RG)
calculations. We find that disorder favors the coating transition,
in the sense that the onset of leaving a Landau--Levich film for a
random substrate with strength $g$ takes place at a dynamic
contact-angle
\begin{equation}
\left.{\theta_{c} \over \theta_e}\right|_l=\alpha_{cl} \left(g
\over \gamma \theta_e^2\right)^{1/3},\label{thetac-l}
\end{equation}
for local dissipation, and
\begin{equation}
\left.{\theta_{c} \over \theta_e}\right|_h={1 \over \sqrt{3}}
+\alpha_{ch} \left(g \over \gamma
\theta_e^2\right)^{2/3},\label{thetac-h}
\end{equation}
for hydrodynamic dissipation, to the leading order. ($\alpha_{cl}$
and $\alpha_{ch}$ are numerical constants to be defined below.)
The value of the roughness exponent at the transition is found to
determine the order of the transition. Although we find that this
exponent acquires non-universal values, it appears that the
predicted nature of the phase transition from the RG calculation
is in agreement with the mean-field results, i.e. second order for
local case and first order for hydrodynamic case, for sufficiently
weak disorder.

For a sufficiently slowly moving contact line, we find that a {\em
pinning transition} takes place, which is the reverse of the
depinning transition, and that it can also be understood as a
roughening transition. We obtain the phase boundary for this
transition corresponding to the advancing and receding contact
angles at the onset of pinning as
\begin{equation}
\left.{\theta_a^2 \over \theta_e^2}\right|_l-1=1-\left.{\theta_r^2
\over \theta_e^2}\right|_l=\alpha_{pl} \left(g \over \gamma
\theta_e^2\right)^{2},     \label{theta-ar-l}
\end{equation}
for the local case, and
\begin{equation}
\left.{\theta_a^2 \over \theta_e^2}\right|_h-1=1-\left.{\theta_r^2
\over \theta_e^2}\right|_h=\alpha_{ph} \left(g \over \gamma
\theta_e^2\right)^{2},     \label{theta-ar-h}
\end{equation}
for the hydrodynamic case, to the leading order, which is in
agreement with a previous prediction by Robbins and Joanny
\cite{RJ1}. ($\alpha_{pl}$ and $\alpha_{ph}$ are numerical
constants to be defined below.) We further find that the roughness
exponent at the pinning threshold is also non-universal.

\begin{figure}
\centerline{\epsfxsize 9.0cm {\epsffile{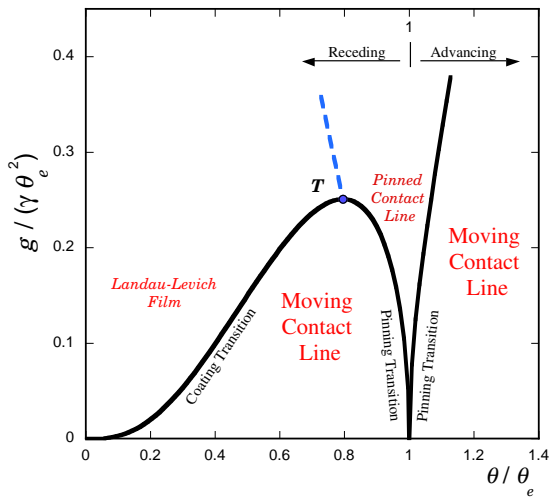}}}
\vskip-3.5truecm \caption{The suggested phase diagram of a contact
line with {\em local} dissipation on a disordered substrate. The
asymptotic forms for the coating and pinning transition lines are
given in Eqs. (\ref{thetac-l}) and (\ref{theta-ar-l}),
respectively.} \label{fig:PD-local}
\end{figure}

\begin{figure}
\centerline{\epsfxsize 8.5cm {\epsffile{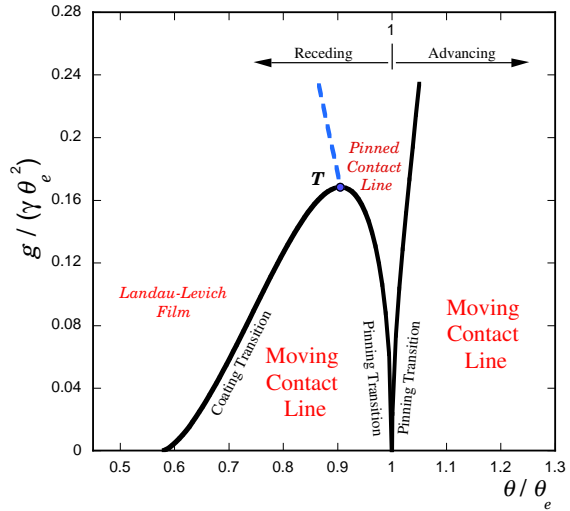}}}
\vskip-3.5truecm \caption{The suggested phase diagram of a contact
line with {\em hydrodynamic} dissipation on a disordered
substrate. The asymptotic forms for the coating and pinning
transition lines are given in Eqs. (\ref{thetac-h}) and
(\ref{theta-ar-h}), respectively. Note that the coating transition
line starts at $1/\sqrt{3} \simeq 0.577$ for zero disorder.}
\label{fig:PD-hydro}
\end{figure}

We combine our results for the coating transition and the pinning
transition, and propose a phase diagram for contact lines with
local dissipation as depicted in Fig.~\ref{fig:PD-local}, and a
corresponding one for contact lines with hydrodynamic dissipation
as depicted in Fig.~\ref{fig:PD-hydro}. In particular, we find
that the phase boundaries corresponding to the coating transition
and the pinning transition meet at a junction point $T$, and
suggest that for sufficiently strong disorder a receding contact
line will leave a Landau--Levich film immediately after depinning.
This corresponds to the dashed lines in Figs.~\ref{fig:PD-local}
and \ref{fig:PD-hydro}. Note that the asymptotic form for the
coating transition lines in Figs.~\ref{fig:PD-local} and
\ref{fig:PD-hydro} are given in Eqs. (\ref{thetac-l}) and
(\ref{thetac-h}), and the asymptotic form for the pinning
transition lines are given in Eqs. (\ref{theta-ar-l}) and
(\ref{theta-ar-h}), respectively.

The rest of this article is organized as follows. In Sec.
\ref{sDynamics}, the main ingredients in the dynamics of the
contact line are discussed, and they are put together in Sec.
\ref{sEquation} where a stochastic dynamical equation is proposed.
In Sec. \ref{sCharacter}, the stochastic dynamical equation is
characterized by its self-affine behavior in terms of various
exponents. The mean-field theory of a moving contact line is
discussed in Sec. \ref{sC-V}, accompanied by a linear relaxation
theory in Sec. \ref{sLinear}. The effect of the nonlinearities is
incorporated using a dynamical RG scheme which is discussed in
Sec. \ref{sNonlinear}, followed by the results in Sec.
\ref{sPhaseDiag} and some discussions in Sec. \ref{sDiscuss}. Some
details related to the differential geometry of the moving liquid
drop is relegated to Appendix \ref{aGeometry}, and finally some
useful asymptotic forms of the results presented in Sec.
\ref{sEquation} are given in Appendix \ref{aCouplings}.

\section{Dynamics of a Deforming Contact Line}  \label{sDynamics}

\begin{figure}
\epsfxsize 7.0cm {\epsffile{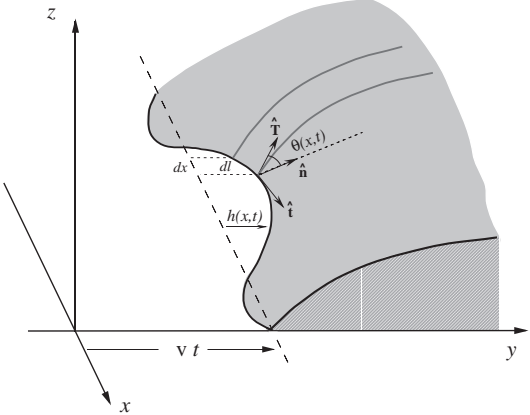}}
\caption{The schematics of the system.} \label{fig:schematics}
\end{figure}

Let us assume that the contact line is oriented along the
$x$-axis, and is moving in the $y$-direction with the position
described by $y(x,t)=v t+h(x,t)$, as depicted in
Fig.~\ref{fig:schematics}.

\subsection{Interfacial Forces}  \label{sInterfacial}

If a line element of length $d l=d x \sqrt{1+(\partial_x h)^2}$ is
displaced by $\delta y(x,t)$, the interfacial energy will be
locally modified by two contributions: (i) the difference between
the solid-vapor $\gamma_{\rm SV}$ and the solid-liquid
$\gamma_{\rm SL}$ interfacial energies times the swept area in
which liquid is replaced by vapor, namely, $(\gamma_{\rm
SV}-\gamma_{\rm SL}) d l \delta y/\sqrt{1+(\partial_x h)^2}$, and
(ii) the work done by the surface tension force, whose direction
is along the unit vector ${\hat {\bf T}}$ that is parallel to the
liquid-vapor interface at the contact and perpendicular to the
contact line, as $\gamma {\hat {\bf T}} \cdot {\hat {\bf y}} d l
\delta y$. Note that we are interested in length scales below the
capillary length, where gravity does not play a role. The overall
change in the interfacial energy of the system can thus be written
as
\begin{equation}
\delta E=\int d x \sqrt{1+(\partial_x h)^2} \left[\frac{\gamma
\cos \theta_e} {\sqrt{1+(\partial_x h)^2}}-\gamma {\hat {\bf T}}
\cdot {\hat {\bf y}} \right]\delta y(x,t), \label{deltaE}
\end{equation}
in which
we have made use of the Young equation. Note that both ``forces''
should be projected onto the $y$-axis when calculating the work
done for a displacement in this direction.

\subsection{Dissipation}  \label{sDissipation}

To calculate the dissipation relevant to the dynamics of the
contact line, we should consider only the {\em normal} component
of the velocity \cite{BR}. If the contact-angle is very small, the
dominant contribution to the dissipation comes from the viscous
losses in the hydrodynamic flows of the liquid wedge \cite{dG1}.
For a slightly deformed contact line, we can assume that the
dissipation is well approximated by the sum of contributions from
wedge-shaped slices with local contact-angles $\theta(x,t)$, as
shown in Fig.~\ref{fig:schematics}. This is a reasonable
approximation because most of the dissipation is taking place in
the singular flows near the tip of the wedge \cite{dG1,dG2}. Using
the result for the dissipation in a perfect wedge which is based
on the lubrication approximation \cite{dG1,Huh}, we can calculate
the dissipation in the hydrodynamic regime as
\begin{equation}
P_{h}={1 \over 2}\int d x \sqrt{1+(\partial_x h)^2} \left(3 \eta
\ell \over \theta(x,t)\right) \left[v+\partial_t h(x,t) \over
\sqrt{1+(\partial_x h)^2}\right]^2,\label{Ph}
\end{equation}
where $\eta$ is the viscosity of the liquid and $\ell$ is a
logarithmic factor of order unity \cite{dG2}. One can show that
the error in the above estimate, which comes from overlap between
the neighboring slices, only leads to curvature terms that are
sub-dominant in the long wavelength limit.

Another physical process that is involved in causing dissipation
is molecular jumps near the tip of the contact line, and is local
in nature \cite{Blake}. Therefore, in any small neighborhood the
amount of dissipation is completely determined by the local value
of the contact line velocity, while all the molecular details of
the dissipation is encoded in an effective friction coefficient
$\mu^{-1}$. The overall dissipation can then be written as
\begin{equation}
P_{l}={1 \over 2 \mu} \int d x \sqrt{1+(\partial_x h)^2}
\left[v+\partial_t h(x,t) \over \sqrt{1+(\partial_x
h)^2}\right]^2. \label{Pl}
\end{equation}

\subsection{Force Balance}  \label{sBalance}

We can find the governing dynamical equation by balancing the
total friction force obtained as $\delta (P_l+P_h)/\delta
\partial_t h(x,t)$ with the interfacial force $-\delta E/\delta h(x,t)$
at each point along the contact line. In the limit of small
contact-angles, we find \cite{small}
\begin{equation}
\left[{1 \over \mu}+{3 \eta \ell \over \theta(x,t)}\right]
{v+\partial_t h(x,t) \over \sqrt{1+(\partial_x h)^2}}={\gamma
\over 2} \left[\theta_e^2-\theta(x,t)^2\right]. \label{D-E}
\end{equation}\noindent
The above equation might simply be recovered by locally applying
the result of Ref.~\cite{BWdG} for straight contact lines, with
the additional geometrical factor (that is needed when the
direction of motion is not perpendicular to the contact line
\cite{BR}) taken into account.

We can introduce a characteristic velocity for the hydrodynamic
friction as $c_{0h}=\gamma \theta_e^3/(3 \eta \ell)$, and a
corresponding velocity for the local friction as $c_{0l}=\mu
\gamma \theta_e^2$. It is then useful to write the dynamical force
balance equation in terms of these characteristic velocities. It
reads
\begin{equation}
\left[{1 \over c_{0l}}+{1 \over c_{0h}}{\theta_e \over
\theta(x,t)}\right] {v+\partial_t h(x,t) \over \sqrt{1+(\partial_x
h)^2}}={1 \over 2} \left[1-{\theta(x,t)^2 \over
\theta_e^2}\right]. \label{D-E2}
\end{equation}\noindent
Note that the relative strength of the two dissipation mechanisms
is characterized by the ratio $c_{0h}/c_{0l}$, and we can readily
obtain the asymptotic form of the equation when local dissipation
is dominant by taking the limit $c_{0h}/c_{0l} \to \infty$, and
the corresponding form when hydrodynamic dissipation dominates by
taking the limit $c_{0h}/c_{0l} \to 0$.

\subsection{Solving for the Surface Profile}  \label{sProfile}

To complete the calculation, we need to solve for the profile of
the surface, and in particular, the angle $\theta(x,t)$ as a
function of $h(x,t)$.

We may assume that the pressure at the surface equilibrates
rapidly enough, so that for the effective study of the long time
dynamics of the contact line it will be sufficient to set the
instantaneous Laplace pressure to zero \cite{JdG1,dG2}. For
sufficiently small contact-angles, the surface profile $z(x,y,t)$
near the contact line can then be found as a solution of the
Laplace equation
\begin{equation}
\nabla^2 z(x,y,t)=0.    \label{Laplace-eqn}
\end{equation}
One can write a general solution for Eq. (\ref{Laplace-eqn}), of
the form
\begin{equation}
z(x,y,t)=\theta (y-v t)+\int {d k \over 2 \pi} \beta(k,t)
\exp\left[i k x-|k| (y-v t)\right],\label{zxy-Laplace}
\end{equation}
that yields the moving liquid wedge profile for a flat contact
line. Imposing the boundary condition $z(x,v t+h(x,t),t)=0$ at the
position of the contact line then yields
\begin{equation}
\beta(k,t)=-\theta \left[h(k,t)+\int {d q \over 2 \pi} |q| h(q,t)
h(k-q,t)+O(h^3)\right].\label{betak-Laplace}
\end{equation}
Note that we have performed the calculation up to the second order
in $h$ to find the leading order nonlinearity in the dynamical
equation, and that since the nonlinear terms neglected in the
expression for Laplace pressure are of the third order in $z$,
this calculation is consistent.

From the slope of the liquid surface at the position of the
contact line (see Fig.~\ref{fig:schematics} and Appendix
\ref{aGeometry}) one can then obtain an expression for the
contact-angle as a function of the contact line deformation. We
find
\begin{eqnarray}
\theta(x,t)&=&\theta \left\{1+b_0 \int {d k \over 2 \pi} |k|
h(k,t)
e^{i k x} \right.\nonumber \\
&+&{1 \over 2} \int {d k \over 2 \pi}{d k' \over 2 \pi} \left[b_1
k k'+b_2 |k| |k'| \right. \nonumber \\
&&\left. +b_3 |k+k'|(|k|+|k'|-|k+k'|)\right] \nonumber \\
&&\left. \times \; h(k,t) h(k',t) e^{i (k+k')
x}\right\},\label{theta(x)}
\end{eqnarray}
with $b_0=1$, $b_1=1$, $b_2=0$, and $b_3=1$. This can then be used
in Eq. (\ref{D-E}) to yield the dynamical equation.

One may, however, question the validity of the instantaneous
pressure relaxation assumption. To improve on this approximation,
one should attempt to solve for the dynamics of the liquid surface
together with the contact line dynamics, and examine the
corresponding time scales for the surface and contact line
relaxations.

This dynamics can be formulated, within the framework of the
lubrication approximation, using a continuity equation of the form
\begin{equation}
\partial_t z(x,y,t)=\nabla \cdot \left({z^3 \over 3 \eta} \nabla
p\right),   \label{continuity}
\end{equation}
where the fluid film is locally described as a Poiseuille flow
under the influence of the gradient of the Laplace pressure
\begin{equation}
p(x,y,t)=- \gamma \nabla^2 z(x,y,t). \label{laplace}
\end{equation}
Combining the above equations then yields
\begin{equation}
\partial_t z(x,y,t)+{\gamma \over 3 \eta}\nabla \cdot \left[z^3 \nabla
\nabla^2 z\right]=0,   \label{lubrication}
\end{equation}
which is the dynamical equation for the surface deformation in the
lubrication approximation \cite{Levich}. To proceed
systematically, one should attempt to solve Eq.
(\ref{lubrication}) for a moving contact line with equilibrium
contact-angle $\theta_e$ subject to the boundary condition $z(x,v
t+h(x,t),t)=0$, perturbatively in powers of the contact line
deformation $h$ up to second order. Unfortunately, this seems to
be a formidable task, because of the complex structure of this
nonlinear partial differential equation. In fact even at the
zeroth order, i.e. for the case of a flat contact line, this
problem is still the subject of much theoretical investigation
\cite{dG1,Voinov,hydro,Huh,lub}.

We can instead try to estimate the surface relaxation time from
Eq. (\ref{lubrication}) using dimensional arguments. If we
consider a deformation of the characteristic size $q^{-1}$ (in
both $x$ and $y$ directions), and put in a wedge-like profile in
the nonlinear term of Eq. (\ref{lubrication}) of the form $z \sim
\theta q^{-1}$, we find that the corresponding (inverse)
relaxation time scales as $\tau^{-1}(q) \sim (\gamma
\theta^3/\eta) q$. Interestingly, this is the same as the time
scale that we find for the contact line relaxation [see Eq.
(\ref{dispersion}) below], and it shows that the instantaneous
surface relaxation assumption is not feasible. However, since
surface relaxation introduces no new time or length scales in the
system, it seems plausible to assume that the contact-angle
profile as a function of the contact line deformation, as obtained
from a full systematic solution of Eq. (\ref{lubrication}), will
still maintain the form given in Eq. (\ref{theta(x)}) in the long
time and long length-scale limit, perhaps with different values
for the numerical coefficients $b_n$. Since we will be interested
only in this limit in the context of the RG calculations, it seems
reasonable to use the general form proposed in Eq.
(\ref{theta(x)}).

\subsection{Disorder}  \label{sDisorder}

In most practical cases, the dynamics of a contact line is
affected by the defects and heterogeneities in the substrate, in
addition to dissipation and elasticity that we have considered so
far. If the interfacial energies $\gamma_{\rm SV}$ and
$\gamma_{\rm SL}$ are space dependent with the corresponding
averages being ${\bar \gamma_{\rm SV}}$ and ${\bar \gamma_{\rm
SL}}$, a displacement $\delta y(x,t)$ of the contact line is going
to lead to a change in energy as
\begin{equation}
\delta E_{d}=\int d x g(x,v t+h(x,t)) \delta y(x,t),\label{dE}
\end{equation}
where
\begin{equation}
g(x,y)=\gamma_{\rm SV}(x,y)-\gamma_{\rm SL}(x,y)-({\bar
\gamma_{\rm SV}}-{\bar \gamma_{\rm SL}}).\label{g-def}
\end{equation}
Incorporating this contribution in the force balance leads to an
extra force term $g(x,v t)$ on the right hand side of
Eq.(\ref{D-E}), which would act as a noise term in the dynamical
equation for contact line deformation of the form
\begin{equation}
\eta(x,t)=\left({\mu \theta \over \theta+3 \eta \mu \ell}\right)
g(x,vt),\label{eta-def}
\end{equation}
to the leading order.

Assuming that the surface disorder has short range correlations
(so that the correlation length is a microscopic length $a$) with
a strength $g$, with a Gaussian distribution described by
\begin{eqnarray}
\langle g(x,y) \rangle&=&0, \nonumber \\
\langle g(x,y) g(x',y')\rangle&=&g^2 a^2\delta(x-x')
\delta(y-y'),\label{g-moments}
\end{eqnarray}
we can deduce the distribution of the noise as
\begin{eqnarray}
\langle \eta(x,t) \rangle&=&0, \nonumber \\
\langle \eta(x,t) \eta(x',t')\rangle &=& 2 D \delta(x-x')
\delta(t-t'),\label{eta-moments}
\end{eqnarray}
where the strength of the noise is given as
\begin{equation}
D={g^2 a^2 \over 2 |v|}\left({\mu \theta \over \theta+3 \eta \mu
\ell}\right)^2.\label{D-def}
\end{equation}

\section{Dynamical Equation of Motion}  \label{sEquation}

We can now put together all the different ingredients of the
dynamics of the contact line that we discussed in Sec.
\ref{sDynamics}, and obtain the governing dynamical equation.
Inserting Eq. (\ref{theta(x)}) in the force balance relation Eq.
(\ref{D-E2}), and adding the noise term of Eq. (\ref{eta-def}), we
find
\begin{eqnarray}
\partial_t h(k,t)&=&-c |k| h(k,t)+\eta(k,t)  \nonumber \\
&&-{1 \over 2} \int {d q \over 2 \pi} \lambda(q,k-q) h(q,t)
h(k-q,t), \label{relax}
\end{eqnarray}
up to the second order in deformations, with a corresponding
nonlinear coupling as
\begin{eqnarray}
\lambda(q,k-q)&=&-\lambda_1 q (k-q)+\lambda_2 |q||k-q| \nonumber \\
&&+\lambda_3 |k|(|q|+|k-q|-|k|).\label{lambda}
\end{eqnarray}
The zeroth order term in the force balance equation provides the
contact-angle--velocity relation as
\begin{equation}
v={c_{0h} \over 2}{\theta \over \theta_e} \left[1-{\theta^2 \over
\theta_e^2} \over 1+{c_{0h} \over c_{0l}}{\theta \over
\theta_e}\right],\label{v-theta-full}
\end{equation}
and we can also find the other coupling constants, namely the
relaxation speed
\begin{equation}
c={c_{0h} \over 2}{\theta \over \theta_e} \left[{\left(3 {\theta^2
\over \theta_e^2}-1+2 {c_{0h} \over c_{0l}}{\theta^3 \over
\theta_e^3}\right) \over \left(1+{c_{0h} \over c_{0l}}{\theta
\over \theta_e}\right)^2} b_0 \right],\label{c-full}
\end{equation}
and the nonlinear coupling constants
\begin{equation}
\lambda_1={c_{0h} \over 2}{\theta \over \theta_e}
\left[{\left({\theta^2 \over \theta_e^2}-1\right) \over
\left(1+{c_{0h} \over c_{0l}}{\theta \over
\theta_e}\right)}+{\left(1-3 {\theta^2 \over \theta_e^2}-2 {c_{0h}
\over c_{0l}}{\theta^3 \over \theta_e^3}\right) \over
\left(1+{c_{0h} \over c_{0l}}{\theta \over \theta_e}\right)^2}
b_1\right],\label{lambda1-full}
\end{equation}
and
\begin{eqnarray}
\lambda_2&=&{c_{0h} \over 2}{\theta \over \theta_e} \left[{\left(3
{\theta^2 \over \theta_e^2}-1+2 {c_{0h} \over c_{0l}}{\theta^3
\over \theta_e^3}\right) \over \left(1+{c_{0h} \over
c_{0l}}{\theta \over \theta_e}\right)^2} b_2 \right. \nonumber \\
&&\left.+{\left(3 {\theta^2 \over \theta_e^2}+{c_{0h} \over
c_{0l}}{\theta \over \theta_e}+3 {c_{0h} \over c_{0l}}{\theta^3
\over \theta_e^3}+{c_{0h}^2 \over c_{0l}^2}{\theta^4 \over
\theta_e^4}\right) \over \left(1+{c_{0h} \over c_{0l}}{\theta
\over \theta_e}\right)^3} (2 b_0^2)\right],\label{lambda2-full}
\end{eqnarray}
and
\begin{equation}
\lambda_3={c_{0h} \over 2}{\theta \over \theta_e} \left[{\left(3
{\theta^2 \over \theta_e^2}-1+2 {c_{0h} \over c_{0l}}{\theta^3
\over \theta_e^3}\right) \over \left(1+{c_{0h} \over
c_{0l}}{\theta \over \theta_e}\right)^2} b_3
\right],\label{lambda3-full}
\end{equation}
in terms of the dynamic contact-angle. The spectrum of the noise
term in Fourier space is characterized as
\begin{eqnarray}
\langle \eta(k,t) \rangle&=&0, \nonumber \\
\langle \eta(k,t) \eta(k',t')\rangle&=&2 D (2 \pi) \delta(k+k')
\delta(t-t'),\label{noise-0}
\end{eqnarray}
where the strength of the noise $D$ is given as in Eq.
(\ref{D-def}), or equivalently as
\begin{equation}
D={a^2 \over \left|1-{\theta^2 \over \theta_e^2}\right|}
{\left(c_{0h} {\theta \over \theta_e}\right) \over \left(1+{c_{0h}
\over c_{0l}}{\theta \over \theta_e}\right)} \left(g \over \gamma
\theta_e^2\right)^2.\label{D-def2}
\end{equation}

The above dynamical equation and its corresponding physical
implications will be discussed in detail in the following
sections. Since we will be attempting to compare the two
alternative dissipation scenarios during these discussions, we
have summarized in Appendix \ref{aCouplings} the limiting form of
the above equations corresponding to each of these mechanisms
separately.

\section{Characterizing the Stochastic Dynamics}  \label{sCharacter}

Due to the presence of the heterogeneities in the substrate, the
contact line undergoes dynamical fluctuations during its (average)
drift motion. These fluctuations, which are governed by the
stochastic dynamical equation given in Sec. \ref{sEquation} [Eq.
(\ref{relax})], can best be characterized by the width of the
contact line, which is defined as
\begin{equation}
W^2(L,t)\equiv {1 \over L^d} \int d^d {\bf x} \langle h({\bf
x},t)^2 \rangle=\int {d^d {\bf k} \over (2 \pi)^d} \langle |h({\bf
k},t)|^2 \rangle,\label{width-def}
\end{equation}
where the averaging is with respect to the noise term in Eq.
(\ref{relax}). Note that we have generalized the contact line to a
$d$-dimensional object of size $L$ so that the dependence on the
dimensionality becomes manifest.

Since the stochastic dynamics described by Eq. (\ref{relax})
corresponds to scale-free fluctuations, the resulting two-point
correlation function should have a scaling form as
\begin{equation}
\langle |h({\bf k},t)|^2 \rangle={1 \over k^{d+2 \zeta}} {\cal
G}(k^z t).\label{two-point}
\end{equation}
The function ${\cal G}(u)$ has the property that it saturates to a
finite value for large $u$, to ensure that a stationary regime can
be achieved in the long time limit. Using the above scaling form,
the width of the contact line will be given as
\begin{equation}
W^2(L,t)\sim \int_{\pi/L}^{\pi/a} {d k \over k^{1+2 \zeta}} {\cal
G}(k^z t),\label{width-1}
\end{equation}
that yields
\begin{equation}
W(L,t)\sim\left\{
\begin{array}{ll} t^{\zeta/z} & ;\; t \ll L^z \\ \\
L^{\zeta} &; \; t \gg L^z  \\
\end{array}\right..\label{WLt}
\end{equation}

From equilibrium phase transitions, say in a magnetic system, we
know that the fluctuations in the overall magnetization is
proportional to the susceptibility, which is a response function.
It is known that this quantity diverges at the critical point for
second order phase transitions, while it stays finite for first
order phase transitions. For a $d$-dimensional system of size $L$,
the divergence appears as $\chi(T_c)\sim M^2/L^d \sim L^{2-\eta}$,
with $\eta$ being a critical exponent.

We can study the overall fluctuations in the order parameter field
for the coating transition $\delta \theta({\bf x},t)=\theta({\bf
x},t)-\theta$ in the contact line problem. One can define
\begin{equation}
{\Theta^2 \over L^d} \equiv {1 \over L^d} \int d^d {\bf x} \; d^d
{\bf x'} \; \langle \delta \theta({\bf x},t) \delta \theta({\bf
x'},t)\rangle,\label{T2-def}
\end{equation}
and use the relation between $\theta({\bf x},t)$ and $h({\bf
x},t)$ given in Eq. (\ref{theta(x)}), to find
\begin{equation}
{\Theta^2 \over L^d} \sim \int d^d {\bf x} \int {d^d {\bf k} \over
(2 \pi)^d} \; k^2 \; e^{i {\bf k} \cdot {\bf x}}\;\langle |h({\bf
k},t)|^2 \rangle,\label{T2-1}
\end{equation}
to the leading order. We can now use the scaling form of Eq.
(\ref{two-point}) in the long time limit, and obtain
\begin{equation}
{\Theta^2 \over L^d} \sim \int d^d {\bf x} \int {d^d {\bf k} \over
(2 \pi)^d} k^2 \; e^{i {\bf k} \cdot {\bf x}}\;{1 \over k^{2
\zeta+d}} \sim \int_a^L {d x \over x^{3-d-2 \zeta}} ,\label{T2-2}
\end{equation}
which yields
\begin{equation}
{\Theta^2 \over L^d}\sim\left\{
\begin{array}{ll} 1 & ; \; \zeta < {2-d \over 2} \\ \\
L^{2 \zeta+d-2} &; \; \zeta > {2-d \over 2}  \\
\end{array}\right..\label{T2L}
\end{equation}
We thus find interestingly that the value of the roughness
exponent at the onset of the coating transition can determine the
order of the dynamical phase transition: the phase transition is
{\em first order} for $\zeta < {2-d \over 2}$, and it is {\em
second order} for $\zeta > {2-d \over 2}$. In the case of a second
order phase transition, we can define a dynamical exponent
\begin{equation}
\eta=4-d-2 \zeta,    \label{eta-exponent}
\end{equation}
based on the above analogy with the equilibrium critical
phenomena, and an order parameter exponent
\begin{equation}
\beta=(1-\zeta) \nu,    \label{beta-exponent}
\end{equation}
from a naive application of the exponent identities, where $\nu$
is the exponent that characterizes the divergence of the
correlation length (see below).

Similarly, we can calculate the magnitude of local order parameter
fluctuations. We find
\begin{equation}
\langle \delta \theta({\bf x},t)^2 \rangle \sim \int {d^d {\bf k}
\over (2 \pi)^d}\; k^2 \langle |h(k,t)|^2
\rangle,\label{theta2-def}
\end{equation}
to the leading order, which yields
\begin{equation}
\langle \delta \theta({\bf x},t)^2 \rangle \sim
\int_{\pi/L}^{\pi/a} {d k \over k^{2 \zeta-1}} {\cal G}(k^z
t).\label{theta2-1}
\end{equation}
and, consequently,
\begin{equation}
\langle \delta \theta({\bf x},t)^2 \rangle\sim\left\{
\begin{array}{ll} t^{2(\zeta-1)/z} & ; \; t \ll L^z \\ \\
\delta \theta_{\rm st}^2(L) &; \; t \gg L^z  \\
\end{array}\right.,\label{theta2Lt}
\end{equation}
where
\begin{equation}
\delta \theta_{\rm st}(L)\sim\left\{
\begin{array}{ll} 1 & ; \; \zeta < 1 \\ \\
\sqrt{\ln(L/a)} & ; \; \zeta=1 \\ \\
L^{\zeta-1} &; \; \zeta > 1  \\
\end{array}\right..\label{theta2L}
\end{equation}
The above result shows that the extent of the order parameter
fluctuations actually depends on the value of the roughness
exponent $\zeta$ in the stationary limit, i.e., it will be finite
for $\zeta < 1$, and unbounded for $\zeta > 1$. It is important to
note that $\zeta > 1$ signals a breakdown of our perturbative
scheme in dealing with the nonlinearities in the system, as one
can show that the neglected nonlinear terms in Eq. (\ref{relax})
can actually be organized as a power series in the parameter
$L^{\zeta-1}$ in the long wavelength limit.

A systematic study of the stochastic dynamics described by Eq.
(\ref{relax}) will yield the values of $z$ and $\zeta$, and thus
provide us with a characterization of the statistical properties
of the moving contact line.

\section{Contact-angle--velocity relation: mean-field theory for
the coating transition}  \label{sC-V}

The dynamics described in Sec. \ref{sEquation}, can be first
studied in the mean-field approximation, where the contact line is
assumed to be a straight line. In this case, the relation between
the dynamic contact-angle of the wetting front, and its velocity
is given by Eq. (\ref{v-theta-full}). It is instructive to examine
the limiting behavior of this equation in the two cases of local
dissipation and hydrodynamic dissipation scenarios, separately.

\subsection{Local Approach}  \label{subB2}

\begin{figure}
\centerline{\epsfxsize 6cm {\epsffile{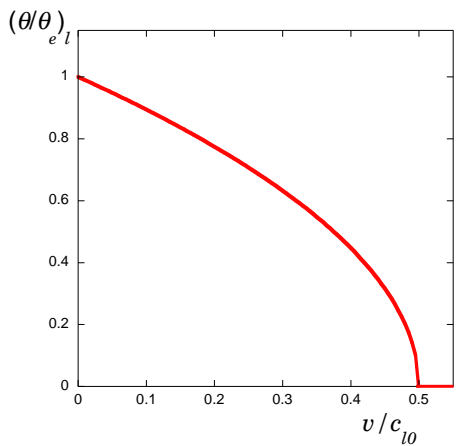}}}
\vskip.3truecm \caption{The reduced order parameter
$(\theta/\theta_e)_l$ as a function of the dimensionless velocity
$v/c_{0l}$ for local mechanism [Eq. (\ref{cav-l})]. The dynamical
phase transition at $v_{cl}/c_{0l}=1/2$ is predicted to be of
second order in this picture.} \label{fig:theta-l}
\end{figure}

In this regime, Eq. (\ref{v-theta-full}) yields [see also
Eq.(\ref{v-theta-l})]
\begin{equation}
\left.{\theta \over \theta_e}\right|_l=\sqrt{1-{2 v \over
c_{0l}}}, \label{cav-l}
\end{equation}
for $v< {c_{0l} \over 2}$, while $\theta=0$ identically for
$v>{c_{0l} \over 2}$. This function is plotted in
Fig.~\ref{fig:theta-l}.

As can be readily seen from Fig.~\ref{fig:theta-l}, increasing $v$
would lead to decreasing values of $\theta$ until at a critical
velocity $v_{cl}={c_{0l} \over 2}$ it finally vanishes
continuously. A vanishing contact-angle presumably corresponds to
formation of a Landau-Levich film. The value of the dynamic
contact-angle $\theta$ serves as the order parameter for this
dynamical phase transition, while $v$ is the tuning parameter. The
continuous vanishing of the order parameter makes the phase
transition classified as second order. As in the general theory of
critical phenomena, a mean-field exponent $\beta=1/2$ is
characterizing the vanishing of the order parameter in terms of
the tuning parameter.

\subsection{Hydrodynamic Approach}  \label{subPGG2}

In this case, Eq.(\ref{v-theta-full}) leads to [see also Eq.
(\ref{v-theta-h})]\footnote{Note that the expression in Eq.
(\ref{cav-h}) is real, and the $i$ is retained only to keep the
appearance of the formula simpler.}
\begin{eqnarray}
\left.{\theta \over \theta_e}\right|_h&=&{1 \over \sqrt{3}}
\left[\left(-{3 \sqrt{3} v \over c_{0h}}-i \sqrt{1-{27 v^2 \over
c_{0h}^2}}\right)^{1/3}\right. \nonumber \\
&&\left.+\left(-{3 \sqrt{3} v \over c_{0h}}+i \sqrt{1-{27 v^2
\over c_{0h}^2}}\right)^{1/3}\right], \label{cav-h}
\end{eqnarray}

\begin{figure}
\centerline{\epsfxsize 6cm {\epsffile{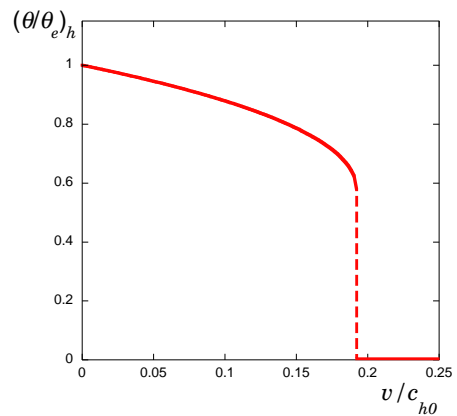}}}
\vskip.3truecm \caption{The reduced order parameter
$(\theta/\theta_e)_h$ as a function of the dimensionless velocity
$v/c_{0h}$ for the hydrodynamic mechanism [Eq. (\ref{cav-h})]. The
dynamical phase transition at $v_{ch}/c_{0h}=1/(3\sqrt{3})\simeq
0.192$ is predicted to be of first order in this picture.}
\label{fig:theta-h}
\end{figure}

The above formula, which holds only for $v< {c_{0h} \over 3
\sqrt{3}}$ has two branches and only the one that recovers
$\theta=\theta_e$ for zero velocity is acceptable as plotted in
Fig.~\ref{fig:theta-h}. While at $v={c_{0h} \over 3 \sqrt{3}}$ we
find $\theta={\theta_e \over \sqrt{3}}$, we expect to have
$\theta=0$ for higher velocities $v>{c_{0h} \over 3 \sqrt{3}}$.
Therefore, the order parameter $\theta$ experiences a finite jump
at the transition velocity $v_{ch}={c_{0h} \over 3 \sqrt{3}}$,
which is the hallmark of a first order phase transition.

\section{Linear Theory}  \label{sLinear}

One can go beyond the simple mean-field treatment, and study the
effect of linear fluctuations on top of the mean-field theory.
There are different aspects to the linear dynamics that one can
elaborate on, as considered in this section.

\subsection{Relaxation}  \label{sRelax}

We can study the relaxation dynamics of a moving contact line only
by using the linear term in Eq. (\ref{relax}). We find that each
deformation mode of wavevector $k$ relaxes with a characteristic
time scale given as
\begin{equation}
\tau^{-1}(k)=c |k|, \label{dispersion}
\end{equation}
where the relaxation velocity $c$ is given as in Eq.
(\ref{c-full}) above. [See also Eqs. (\ref{c-l}) and (\ref{c-h})
for the limiting forms.]

In fact, one can show from Eqs. (\ref{v-theta-full}) and
(\ref{c-full}), that the relaxation velocity $c$ is related to the
slope of the contact-angle--velocity curve as
\begin{equation}
c=-b_0 \theta {d v \over d \theta}. \label{cdv/dtheta}
\end{equation}
This proves that the onset of instability in the contact-angle,
signalled by a diverging slope of $d \theta/d v$, exactly
corresponds to where the relaxation velocity $c$ vanishes. In
other words, exactly at the onset of the coating transition (as
found in the mean-field scheme of Sec. \ref{sC-V}), where the
order parameter has a singular behavior, the relaxation becomes
infinitely slow, and a distorted the contact line does not relax.

This ``coincidence'' suggests strongly that the coating
transition---the onset of leaving a Landau--Levich film---can be
described as a dynamical phase transition in terms of the
deformations of the contact line and its statistical roughness on
a disordered substrate.

\subsection{Fluctuations}  \label{sFluct}

To account for the statistical fluctuations of the moving contact
line on a disordered substrate, we can also include the noise term
in Eq. (\ref{relax}), and calculate the width of the contact line.
We find
\begin{equation}
W(L,t) \sim \left\{\begin{array}{ll} \sqrt{t}, &\; t \ll {a \over
c},  \\
\sqrt{\ln\left[c t \over a\right]}, &\; {a \over c} \ll t \ll {L
\over
c},  \\
\sqrt{\ln\left(L \over a\right)}, &\; t \gg {L \over c}.
\end{array} \right. \label{W1}
\end{equation}
There are two important time scales in the above equation: (i) the
``microscopic'' time $a/c$ that corresponds to the crossover from
local diffusive dynamics to collective motion along the contact
line, and (ii) the ``macroscopic'' time $L/c$ that corresponds to
the crossover to the stationary state. We thus find from Eq.
(\ref{W1}) that the stochastic deformations of the contact line
are characterized by $\zeta_0=0$ and $z_0=1$, within the linear
theory.

\subsection{Breakdown}  \label{sBreak}

The nonlinear term in Eq. (\ref{relax}) will modify the above
results if it becomes appreciable at long length scales, as
compared to the linear term. To examine under what conditions this
may take place, we can estimate the ratio of the two terms in Eq.
(\ref{relax}), which scales as
$$
{c_0 (L/a)^{2 \zeta_0-1} \over c (L/a)^{\zeta_0}} \sim {a c_0
\sqrt{\ln\left(L/a\right)}\over L c},
$$
where $c_0$ could be set by either $c_{0l}$ or $c_{0h}$. The
nonlinear term is thus appreciable {\it only} when the smallest
time scale in the linear theory $a/c$ becomes comparable to
$L/c_0$, which happens near the coating transition when $c$
becomes small (see Sec. \ref{sRelax}). This confirms that a
consistent study of the coating transition should take a proper
account of the nonlinear terms.

\section{Effect of the Nonlinear Term}  \label{sNonlinear}

Let us now attempt to systematically study the dynamical phase
transition in the moving contact line using an RG scheme. The
dynamical equation, which can be generally written in $d$
dimensions as \cite{notation}
\begin{eqnarray}
\partial_t h({\bf k},t)&=&-c k \; h({\bf k},t)+\eta({\bf k},t) \nonumber \\
&-&{1 \over 2} \int {d^d {\bf q} \over (2 \pi)^d} \lambda({\bf
q},{\bf k}-{\bf q}) h({\bf q},t) h({\bf k}-{\bf
q},t),\label{KPZ-RG1}
\end{eqnarray}
with $\lambda({\bf q},{\bf k}-{\bf q})$ given as in Eq.
(\ref{lambda}), belongs to the general class of
Kardar-Parisi-Zhang (KPZ) equations
\cite{KPZ,Medina,Halpin,Frey,Kay,Venkat}. We take a noise spectrum
given by
\begin{eqnarray}
\langle \eta({\bf k},t) \rangle&=&0, \nonumber \\
\langle \eta({\bf k},t) \eta({\bf k}',t')\rangle&=&2 D (2 \pi)^d
\delta^d({\bf k}+{\bf k}') \delta(t-t'),\label{noise-1}
\end{eqnarray}
and employ standard RG techniques following Ref. \cite{Medina} to
calculate the RG equations describing the flow of the coupling
constants.

\subsection{Perturbation Theory}  \label{sPT}

To perform this calculation, we first need to construct a
perturbation theory that takes into account the nonlinear term.
This is done more easily if we perform a Fourier transformation in
the time variable in Eq. (\ref{KPZ-RG1}), which yields
\begin{eqnarray}
-i \omega h({\bf k},\omega)&=&-c k \; h({\bf k},\omega)+\eta({\bf k},\omega) \nonumber \\
&-&{1 \over 2} \int {d \Omega \over 2 \pi} {d^d {\bf q} \over (2
\pi)^d}
\lambda({\bf q},{\bf k}-{\bf q}) \nonumber \\
&&\times \; h({\bf q},\Omega) h({\bf k}-{\bf
q},\omega-\Omega).\label{KPZ-RG2}
\end{eqnarray}
This equation can then be re-written in the form
\begin{eqnarray}
h({\bf k},\omega)&=& G_0({\bf k},\omega) \eta({\bf k},\omega) \nonumber \\
&-&{1 \over 2} G_0({\bf k},\omega) \int {d \Omega \over 2 \pi}
{d^d {\bf q}
\over (2 \pi)^d} \lambda({\bf q},{\bf k}-{\bf q}) \nonumber \\
&&\times \; h({\bf q},\Omega) h({\bf k}-{\bf
q},\omega-\Omega),\label{KPZ-RG3}
\end{eqnarray}
in which the bare response function is given as
\begin{equation}
G_0({\bf k},\omega)={1 \over c k-i \omega}. \label{G_0kw}
\end{equation}
We can then define the full response function $G({\bf k},\omega)$
via
\begin{equation}
h({\bf k},\omega)=G({\bf k},\omega) \eta({\bf k},\omega),
\label{Gkw}
\end{equation}
and use the spectrum for the Fourier transform of the noise as
\begin{eqnarray}
\langle \eta({\bf k},\omega) \rangle&=&0, \nonumber \\
\langle \eta({\bf k},\omega) \eta({\bf k}',\omega')\rangle&=&2 D
(2 \pi)^{d+1} \delta^d({\bf k}+{\bf k}')
\delta(\omega+\omega'),\label{noise-2}
\end{eqnarray}
to find the response function perturbatively in the
$\lambda$-parameters.

Up to second order in the perturbation theory, we find
\begin{eqnarray}
G({\bf k},\omega)&=& G_0({\bf k},\omega)+4 \times {1 \over 4}
G_0({\bf k},\omega)^2
\times 2 D \nonumber \\
&& \times \; \int {d \Omega \over 2 \pi} {d^d {\bf q}
\over (2 \pi)^d} \lambda({\bf q},{\bf k}-{\bf q}) \lambda(-{\bf q},{\bf k})\nonumber \\
&&\times \; G_0({\bf q},\Omega) G_0(-{\bf q},-\Omega)G_0({\bf
k}-{\bf q},\omega-\Omega),\label{GG-0-RG1}
\end{eqnarray}
which can be re-written as
\end{multicols}
\begin{eqnarray}
G^{-1}({\bf k},\omega)&=& G_0^{-1}({\bf k},\omega)-2 D
\int_{-\infty}^{\infty} {d \Omega \over 2 \pi} \int_{0}^{\Lambda}
q^{d-1} d q \; {S_{d-1}\over (2 \pi)^d}
\int_{0}^{\pi} d \theta \sin^{d-2} \theta \nonumber \\
&\times& \; {1 \over \left[c \sqrt{q^2+k^2-2 k q \cos \theta}-i
(\omega-\Omega)\right]
(c^2 q^2+\Omega^2)}\nonumber \\
&\times& \; \left[\lambda_1 (q^2-q k \cos \theta) +\lambda_2 q
\sqrt{q^2+k^2-2 k q \cos \theta}+\lambda_3 k q
+\lambda_3 k \sqrt{q^2+k^2-2 k q \cos \theta}-\lambda_3 k^2  \right]\nonumber \\
&\times& \; \left[\lambda_1 k q \cos \theta+\lambda_2 q
k+\lambda_3 q \sqrt{q^2+k^2-2 k q \cos \theta}+\lambda_3 k
\sqrt{q^2+k^2-2 k q \cos \theta}-\lambda_3 (q^2+k^2-2 k q \cos
\theta) \right],\label{GG-0-RG2}
\end{eqnarray}
\begin{multicols}{2}\noindent
where $\Lambda=\pi/a$ is an upper cutoff for the wavevector set by
an inverse microscopic length scale, and $S_d$ is the surface area
of a unit sphere in $d$ dimensions. We can then perform the
frequency integration, expand the integrand in powers of $k/q$
(that is justified because we are interested in the long
wavelength behavior of the system), and integrate over the angular
variable, to obtain
\begin{eqnarray}
G^{-1}({\bf k},\omega)&=& G_0^{-1}({\bf k},\omega)-{D \over 2 c^2}
{S_{d}\over (2 \pi)^d} \nonumber \\
&&\times \; (\lambda_1+\lambda_2)(\lambda_2+\lambda_3) \left(\int
q^{d} d q \right) \; k .\label{GG-0-RG3}
\end{eqnarray}
Assuming a same form for the full response function as the bare
one, namely
\begin{equation}
G({\bf k},\omega)={1 \over c_R k-i \omega}, \label{G_kw}
\end{equation}
one then obtain a renormalized elastic constant as
\begin{equation}
c_R=c \left\{1- \left[(\lambda_1+\lambda_2)(\lambda_2+\lambda_3) D
\over 2 c^3 \right]{S_{d}\over (2 \pi)^d} \left(\int q^{d} d q
\right)\right\}.\label{sigma-R}
\end{equation}

A similar calculation can be performed to obtain the renormalized
noise amplitude, that is defined via
\begin{equation}
\langle h({\bf k},\omega) h(-{\bf k},-\omega)\rangle=2 D_R G({\bf
k},\omega)G(-{\bf k},-\omega).\label{D-R-def}
\end{equation}
one obtains
\begin{eqnarray}
2 D_R&=&2 D+2 \times \left(1 \over 2\right)^2 \times (2 D)^2 \int
{d \Omega \over 2 \pi} {d^d {\bf q}
\over (2 \pi)^d} \nonumber \\
&&\times \; \lambda({\bf q},{\bf k}-{\bf q}) \lambda(-{\bf q},{\bf
q}-{\bf k}) G_0({\bf q},\Omega)
G_0(-{\bf q},-\Omega)\nonumber \\
&&\times \; G_0({\bf k}-{\bf q},\omega-\Omega)G_0({\bf q}-{\bf
k},\Omega-\omega),\label{D-RG1}
\end{eqnarray}
which in the small $k$ limit yields
\begin{equation}
D_R=D \left\{1+ \left[(\lambda_1+\lambda_2)^2 D \over 4 c^3
\right]{S_{d}\over (2 \pi)^d} \left(\int q^{d} d q
\right)\right\}.\label{D-R}
\end{equation}

Finally, one can show that similar to original KPZ problem
\cite{KPZ}, none of the $\lambda$-parameters are renormalized, so
that we have
\begin{equation}
\lambda_R({\bf q},{\bf k}-{\bf q})=\lambda({\bf q},{\bf k}-{\bf
q}).\label{lambda-R}
\end{equation}
We can now use the results of the perturbation theory, to
construct a perturbative RG scheme.

\subsection{Renormalization Group Calculation}  \label{sRG}

In order to recapitulate the perturbation theory into an RG
calculation, we should only integrate out over a layer of
wavevectors from $\Lambda/b$ to $\Lambda$ an see how the coupling
constants are affected by that. This step, which makes up the
coarse graining procedure, will lead to the same results as in
Eqs. (\ref{sigma-R}), (\ref{D-R}), and (\ref{lambda-R}), in which
the wavevector integration reads $\int_{\Lambda/b}^{\Lambda} q^{d}
d q={\Lambda^{d+1} \over d+1}\left(1-b^{-(d+1)}\right)$. This
should then be followed by the scale transformations $x \to b x$,
$t \to b^z t$, and $h(x,t) \to b^\zeta h(x,t)$, which for the
scale factor of the form $b=e^l$ and for infinitesimal values of
$l$ yields the following RG flow equations for the coupling
constant:
\begin{eqnarray}
&&{d c \over d l}= c \left[z-1-U\right], \nonumber  \\
&&{d \lambda({\bf q},{\bf k}-{\bf q}) \over d l}= \lambda({\bf
q},{\bf k}-{\bf q}) \left(\zeta+z-2\right),
\label{RG} \\
&&{d D \over d l}= D \left[z-2 \zeta-d
+\left({\lambda_1+\lambda_2 \over \lambda_2+\lambda_3}\right)
{U \over 2}\right], \nonumber
\end{eqnarray}
in which
\begin{equation}
U={\pi S_d (\lambda_1+\lambda_2) (\lambda_2+\lambda_3) D \over (2
a)^{d+1} c^3}.\label{Udef}
\end{equation}

To study the fixed point structure of the above set of flow
equations, we set $z=1+U$ and $\zeta=1-U$, and look at the flow
equation for $U$:
\begin{equation}
{d U \over d l}=-(d+1) U+\left[6+\left({\lambda_1+\lambda_2 \over
\lambda_2+\lambda_3}\right)\right] {U^2 \over 2}, \label{U-RGflow}
\end{equation}
which has two stable fixed points at $U=0$ (linear theory) and
$U=\infty$ (strong coupling), as well as an intermediate unstable
fixed point at
\begin{equation}
U=U^*\equiv{2(d+1) \over
6+(\lambda_1+\lambda_2)/(\lambda_2+\lambda_3)}.\label{U*}
\end{equation}
For $U<U^*$, the nonlinearity is irrelevant and the exponents are
given by the linear theory, i.e. $\zeta_0=0$ and $z_0=1$, while
for $U>U^*$ the behavior of the system is governed by a strong
coupling fixed point which cannot be studied perturbatively. The
fixed point at $U^*$ corresponds to a roughening transition of the
moving contact line. The exponents at the transition are
\begin{eqnarray}
z&=&1+U^*, \nonumber \\
\zeta&=&1-U^*,\label{z-zeta}
\end{eqnarray}
which turn out to be nonuniversal. The roughening transition
corresponds to the limit of stability of the moving contact line
phase. The phase described by the strong coupling fixed point
could presumably correspond to a Landau--Levich film or a pinned
contact line.

We can also study how the transition is approached by linearizing
the flow equation near the fixed point. Setting $U=U^*+\delta U$,
we find $d \delta U/dl=(d+1) \delta U$ that would imply divergence
of the correlation length near the transition as
\begin{equation}
\xi \sim |\delta U|^{-\nu},\label{xi}
\end{equation}
with
\begin{equation}
\nu={1 \over d+1}.\label{nu}
\end{equation}
The correlation length corresponds to the typical size of rough
segments in the contact line, which should diverge as the
transition is approached (See Fig.~\ref{fig:xi}).

\section{Fixed Point Equation: Phase diagram and Exponents}  \label{sPhaseDiag}

In principle, the position of a phase boundary that separates the
different phases could be obtained from a fixed point equation
[such as Eq. (\ref{U*})]. However, in most RG studies the relation
between the phenomenological parameters in the theory and the
microscopic parameters are not known, and thus the fixed point
equation cannot help us obtain the phase diagram of the system in
terms of the real control parameters.

In the present case, however, the fact that we have used some
physical models to arrive at the the dynamical equations allows us
to make such direct connections. If we use the relations obtained
for the parameters as a function of the contact-angle in Sec.
\ref{sEquation} (and Appendix \ref{aCouplings}), and insert them
in the fixed point equation Eq. (\ref{U*}) with $d=1$, which can
be written as
\begin{equation}
{D \over a^2}={8 \over \pi} {c^3 \over
(\lambda_1+\lambda_2)(\lambda_1+7 \lambda_2+6 \lambda_3)},
\label{fixed-point-2}
\end{equation}
we can map out the phase diagram of the system, and calculate the
value of the exponents.

In the local case, using Eqs. (\ref{v-theta-l}-\ref{D-l}) in Eq.
(\ref{fixed-point-2}) yields an equation for the phase boundary as
\end{multicols}
\begin{equation}
\left.{g \over \gamma \theta_e^2}\right|_l={4 \left({2 b_0^3 \over
\pi}\right)^{1/2} \left({\theta \over \theta_e}\right)^3 \left|1-
{\theta^2 \over \theta_e^2}\right|^{1/2}\over \left[1-(1-2 b_1+2
b_2+2 b_0^2){\theta^2 \over \theta_e^2}\right]^{1/2} \left[1-(1-2
b_1+14 b_2+14 b_0^2+12 b_3){\theta^2 \over \theta_e^2}
\right]^{1/2}}. \label{g-theta-l}
\end{equation}
Similarly, for the hydrodynamic case, using Eqs.
(\ref{v-theta-h}-\ref{D-h}) in Eq. (\ref{fixed-point-2}) yields an
equation for the phase boundary as
\begin{equation}
\left.{g \over \gamma \theta_e^2}\right|_h={6 \left({3 b_0^3 \over
\pi}\right)^{1/2} \left({\theta^2 \over \theta_e^2}-{1 \over 3
}\right)^{3/2} \left|1- {\theta^2 \over
\theta_e^2}\right|^{1/2}\over \left[(b_1-b_2-1)+(1-3 b_1+3 b_2+6
b_0^2){\theta^2 \over \theta_e^2}\right]^{1/2} \left[(b_1-7 b_2-6
b_3-1)+(1-3 b_1+21 b_2+42 b_0^2+18 b_3){\theta^2 \over \theta_e^2}
\right]^{1/2}}, \label{g-theta-h}
\end{equation}
These phase boundaries are plotted in Figs.~\ref{fig:PD-local} and
\ref{fig:PD-hydro} (solid line) for a choice of parameters
$b_0=1$, $b_1=1$, $b_2=-1$, and $b_3=-1$.

The phase boundary that corresponds to a roughening transition of
the moving contact line has two different branches. The first
branch that happens at relatively high velocities, presumably
corresponds to the onset of leaving a Landau--Levich film. In the
local case, the phase boundary for this transition start at zero
contact-angle for zero disorder, and has a limiting form as
reported in Eq. (\ref{thetac-l}) above, with
\begin{equation}
\alpha_{cl}={(2 \pi)^{1/6} \over 2 \sqrt{b_0}}, \label{alpha-c-l}
\end{equation}
whereas in the hydrodynamic case the boundary starts at a finite
value of the contact-angle, with the limiting form as reported in
Eq. (\ref{thetac-h}), in which
\begin{equation}
\alpha_{ch}={1 \over 2 \sqrt{3} b_0} \left[{\pi \over 6} (3
b_0^2-1)(21 b_0^2-1) \right]^{1/3}. \label{alpha-c-h}
\end{equation}
We can calculate the value of the exponents along these phase
boundaries, and as it turns out they are non-universal. We find
\begin{eqnarray}
\zeta_{cl}&=&1+2 (2 \pi)^{1/3} \left(b_0^2+b_2+b_3 \over b_0
\right)
\left(g \over \gamma \theta_e^2\right)^{2/3},   \label{zeta-LL-l} \\
\nonumber \\
z_{cl}&=&1-2 (2 \pi)^{1/3} \left(b_0^2+b_2+b_3 \over b_0 \right)
\left(g \over \gamma \theta_e^2\right)^{2/3}, \label{z-LL-l}
\end{eqnarray}
for the local case, and
\begin{eqnarray}
\zeta_{ch}&=&{3-1/(3 b_0^2) \over 7-1/(3 b_0^2)}-{(36 \pi)^{1/3}
\over b_0} {(3 b_0^2-1)^{1/3} \over (21 b_0^2-1)^{5/3}} \left[3
b_0^2(b_1+b_3-1)-b_2-b_3 \right]
\left(g \over \gamma \theta_e^2\right)^{2/3},   \label{zeta-h} \\
\nonumber \\
z_{ch}&=&{11-1/(3 b_0^2) \over 7-1/(3 b_0^2)}+{(36 \pi)^{1/3}
\over b_0} {(3 b_0^2-1)^{1/3} \over (21 b_0^2-1)^{5/3}} \left[3
b_0^2(b_1+b_3-1)-b_2-b_3 \right] \left(g \over \gamma
\theta_e^2\right)^{2/3},   \label{z-h}
\end{eqnarray}
\begin{multicols}{2}\noindent
for the hydrodynamic case.

The other branch of the phase boundaries appear very near the
equilibrium contact-angle corresponding to low velocities, and
presumably corresponds to the onset of pinning. In other words,
the pinning transition (as opposed to the depinning transition)
can be described as a roughening of the contact line imposed by
the static minimum energy configuration on the disordered
substrate, as it slows down to rest. The shape of the phase
boundaries for weak disorder can be found for the local case as in
Eq. (\ref{theta-ar-l}) above, with
\begin{equation}
\alpha_{pl}={3 \pi \over 4 b_0^3} (b_0^2+b_2-b_1)^2 \left[{1 \over
6}+\left(b_0^2+b_2+b_3 \over b_0^2+b_2-b_1\right)\right],
\label{alpha-p-l}
\end{equation}
and, for the hydrodynamic case as in Eq. (\ref{theta-ar-h}) above,
with
\begin{equation}
\alpha_{ph}={3 \pi \over 4 b_0^3} (3 b_0^2+b_2-b_1)^2 \left[{1
\over 6}+\left(3 b_0^2+b_2+b_3 \over 3
b_0^2+b_2-b_1\right)\right]. \label{alpha-p-h}
\end{equation}
The above prediction for the phase boundary of the pinning
transition is in agreement with a previous prediction based on the
hysteresis in receding and advancing contact-angles \cite{RJ1}. We
can calculate the corresponding exponents along the depinning
transition phase boundary, and find
\begin{eqnarray}
\zeta_{pl}&=&{1+{1 \over 2}\left(b_0^2+b_2-b_1 \over
b_0^2+b_2+b_3\right) \over 3+{1 \over 2}\left(b_0^2+b_2-b_1 \over
b_0^2+b_2+b_3\right)},   \label{zeta-d-l} \\
\nonumber \\
z_{pl}&=&{5+{1 \over 2}\left(b_0^2+b_2-b_1 \over
b_0^2+b_2+b_3\right) \over 3+{1 \over 2}\left(b_0^2+b_2-b_1 \over
b_0^2+b_2+b_3\right)}, \label{z-d-l}
\end{eqnarray}
for the local case, and
\begin{eqnarray}
\zeta_{ph}&=&{1+{1 \over 2}\left(3 b_0^2+b_2-b_1 \over 3
b_0^2+b_2+b_3\right) \over 3+{1 \over 2}\left(3
b_0^2+b_2-b_1 \over 3 b_0^2+b_2+b_3\right)},   \label{zeta-d-h} \\
\nonumber \\
z_{ph}&=&{5+{1 \over 2}\left(3 b_0^2+b_2-b_1 \over 3
b_0^2+b_2+b_3\right) \over 3+{1 \over 2}\left(3 b_0^2+b_2-b_1
\over 3 b_0^2+b_2+b_3\right)}, \label{z-d-h}
\end{eqnarray}
for the hydrodynamic case.

The two branches of the phase boundaries meet at a junction point,
called $T$ in Figs.~\ref{fig:PD-local} and \ref{fig:PD-hydro},
where they both develop a vanishing slope as obtained from Eqs.
(\ref{g-theta-l}) and (\ref{g-theta-h}).

Although the fact that we do not know the values of the numerical
constants $b_n$ makes us unable to predict the values of the
exponents, we can nevertheless put some bounds on them using
simple physical requirements. For example, we expect on physical
grounds that the critical contact-angle for the coating transition
increases with disorder. Then Eqs. (\ref{thetac-h}) and
(\ref{alpha-c-h}), together with a requirement that we have a
roughness exponent that is less than one, imply that $b_0^2> 1/3$,
which results in a criterion
\begin{equation}
{1 \over 3}< \zeta_{ch}  <{3 \over 7},\label{ineq-zeta-LL-h}
\end{equation}
for the roughness exponent in the hydrodynamic case. In the local
case, to have a roughness exponent that is less than one we should
have $b_0^2+b_2+b_3 < 0$, as can be seen from Eq.
(\ref{zeta-LL-l}).

From the above results and bounds on the values of the roughness
exponents in the two different approaches, and the analysis of
Sec. \ref{sCharacter}, we can conclude that at least in the weak
disorder limit $\zeta_{cl}  > {1 \over 2}$ and $\zeta_{ch}  < {1
\over 2}$, and thus the coating transition remains to be first
order in the hydrodynamic picture and second order in the local
picture even beyond mean-field theory, i.e. within the RG scheme.

On the pinning transition boundary a similar requirement that the
advancing (receding) contact-angle increases (decreases) with
disorder (together with a requirement that we have a roughness
exponent that is less than one) leads to $\left(b_0^2+b_2-b_1
\over b_0^2+b_2+b_3\right)>0$, as can be seen from Eqs.
(\ref{theta-ar-l}) and (\ref{alpha-p-l}), which implies
\begin{equation}
\zeta_{pl} > {1 \over 3},\label{ineq-zeta-d-l}
\end{equation}
and $\left(3 b_0^2+b_2-b_1 \over 3 b_0^2+b_2+b_3\right)>0$, as can
be seen from Eqs. (\ref{theta-ar-h}) and (\ref{alpha-p-h}), which
implies
\begin{equation}
\zeta_{ph} > {1 \over 3}.\label{ineq-zeta-d-h}
\end{equation}
The above results are in agreement with the experimentally
observed roughness exponent of about $0.5$ for both liquid Helium
on a Cesium substrate \cite{Rolley2} and water on glass
experiments \cite{Rolley3}.

\section{Discussion}  \label{sDiscuss}

The above study of the dynamics of a moving contact line on a
disordered substrate reveals that both the instability
corresponding to the onset of leaving a Landau--Levich film at
high velocities, and the pinning of the contact line to the
substrate at low velocities, can be described in a unified
framework as roughening transitions in the contact line. For
advancing contact lines, the phase boundary corresponding to
pinning extends continuously to the strong disorder regime. For
the receding case, however, the phase boundaries corresponding to
pinning and to Landau--Levich transition asymptote to a maximum
with zero slope, where we identify as a junction point, denoted by
$T$ in Figs.~\ref{fig:PD-local} and \ref{fig:PD-hydro}.

For stronger disorder, the pinned contact line presumably goes
directly into the Landau--Levich phase: the edge of the liquid
drop is pinned so strongly that it remains still when the liquid
wedge starts to move upon decreasing the contact-angle, and thus a
film is left behind. In this sense, at the dashed line in the
phase diagram (Figs.~\ref{fig:PD-local} and \ref{fig:PD-hydro})
nothing really happens to the contact line itself, while the body
of the liquid drop is ``depinned.'' This behavior has in fact been
observed in experiments with liquid Helium on a Cesium substrate,
where the receding contact-angle has always been found to vanish
identically immediately after depinning \cite{Rolley,Rolley2}. It
is also interesting to note that a non-universality of the
roughness exponent at the depinning threshold, manifesting itself
in the form of a dependence on the value of the contact-angle, has
also been reported in Ref. \cite{Rolley2} .

One can get a plausible picture of the roughening transition in
terms of fluctuating domains of different sizes, in analogy with
equilibrium phase separations in, say, binary mixtures. As it is
sketched in Fig.~\ref{fig:xi}, a portion of the moving contact
line can be instantaneously pinned to a defect on the substrate,
thereby nucleating a domain where the liquid is pinned to the
substrate. These domains are rough, because the contact line has
to conform to the minimal energy configuration imposed by the
substrate disorder, and they can exist in length scales up to the
correlation length $\xi$. As the transition is approached, the
correlation length increases until at some point one of the
domains enlarge to a macroscopic scale and span the whole system,
corresponding to the divergence of the correlation length at the
critical point. Interestingly, the same picture can be attributed
to both the coating transition and the pinning transition. The
difference comes from the fact that in the pinning transition the
substrate can stop both the contact line and the liquid from
motion, while in the coating transition where the liquid is
already moving at relatively high velocities it can only stop the
contact line and not the liquid, as mentioned above.

One can use a scaling argument to account for the power law in
Eqs. (\ref{thetac-l}) and (\ref{thetac-h}). If we take the above
KPZ equation and make the scale transformations $t \to c t$, and
$h \to \sqrt{D/c} h$, the coefficient of the linear term as well
as the strength of the noise term will be set to one, and the only
remaining coupling constant (the coefficient of the nonlinear
term) will have the form $\lambda D^{1/2}/c^{3/2}$. We expect the
roughening transition to take place when this coupling constant is
of order unity, which yields Eqs. (\ref{thetac-l}) and
(\ref{thetac-l}) when the values $D \sim g^2$, $c_l \sim
{\theta_c^2 \over \theta_e^2}$, $c_h \sim \left({\theta_c \over
\theta_e}-{1 \over \sqrt{3}}\right)$, and $\lambda \sim {\rm
const.}$ are used.

In treating the disorder contribution, we have made the assumption
that the dependence in the noise term $\eta$ on the shape of the
contact line can be neglected for any non-vanishing drift
velocity. While this approximation can be supported by a naive
power counting, it may be argued that more a sophisticated
functional RG approach is necessary to deal with this dependence
\cite{Ertas,LeDoussal}, since upon approaching the pinning
transition the contact line velocity becomes progressively small.
In this sense, the exact values of the exponents derived for the
pinning transition boundary, which are non-universal anyway, and
the predictions about the nature of the junction point may not be
trusted. However, we do believe that the main features of the
above results---the non-universality of the exponents, the
existence of the junction point, and the form of the pinning phase
boundary, will persist.

We finally mention that this work can hopefully motivate two types
of experiments. One can try to probe the relaxation of a moving
contact line, similar to the Ondarcuhu--Veyssie experiment on a
static contact line \cite{exp1}, and measure the velocity
dependence of the dispersion relation. This dependence could be
used to determine the dominant dissipation mechanism
\cite{RE-PRE}. It is also interesting to systematically study the
coating transition (onset of leaving a Landau--Levich film) for
receding contact lines on a disordered substrate using video
microscopy techniques. In particular, it would be interesting to
look for a roughening of the contact line before the
Landau--Levich film is formed, and measure the corresponding
roughness exponents.

\acknowledgments

We are grateful to A. Ajdari, J. Bico, R. Bruinsma, P.G. de
Gennes, C. Guthmann, M. Kardar, W. Krauth, L. Limat, S. Moulinet,
D. Qu\'er\'e, E. Rolley, A. Rosso, and H. Stone for valuable
discussions and comments. One of us (R.G.) would like to thank the
group of Prof. de Gennes at College de France for their
hospitality and support during his visit.

\end{multicols}
\appendix

\section{Geometry of the Surface near the Contact Line}  \label{aGeometry}

Here we briefly review some of the aspects of the differential
geometry of surfaces that are useful in this context. The free
surface of the liquid can be described by the embedding
\begin{equation}
{\bf R}(x,y)=(x,y,z(x,y)), \label{Rxyz}
\end{equation}
using which we can find two independent unit tangent vectors to
the surface at each point
\begin{equation}
{\hat {\bf t}}_x \equiv {\partial_x {\bf R} \over |\partial_x {\bf
R}|}={1 \over \sqrt{1+(\partial_x z)^2}}(1,0,\partial_x
z),\label{tx}
\end{equation}
and
\begin{equation}
{\hat {\bf t}}_y \equiv {\partial_y {\bf R} \over |\partial_y {\bf
R}|}={1 \over \sqrt{1+(\partial_y z)^2}}(0,1,\partial_y
z).\label{ty}
\end{equation}
We can also define the unit tangent vector for the contact line
(see Fig.~\ref{fig:schematics}) as
\begin{equation}
{\hat {\bf t}}={1 \over \sqrt{1+(\partial_x h)^2}}(1,\partial_x
h,0).\label{t}
\end{equation}
To find the unit vector ${\hat {\bf T}}$, that shows the local
direction at which the surface tension force is acting (see
Fig.~\ref{fig:schematics}), we should note that it is tangent to
the surface, so it can be written as
\begin{equation}
{\hat {\bf T}}=u_x {\hat {\bf t}}_x+u_y {\hat {\bf
t}}_y.\label{Ttxty}
\end{equation}
Imposing the requirements that it is perpendicular to the contact
line ${\hat {\bf T}} \cdot {\hat {\bf t}}=0$ and that it is a unit
vector ${\hat {\bf T}}^2=1$, yields the two parameters, and we
obtain
\begin{equation}
{\hat {\bf T}}={(-\partial_x h,1,-\partial_x h \partial_x
z+\partial_y z)\over \left\{1+(\partial_y z)^2+(\partial_x h )^2
\left[1+(\partial_x z)^2\right]-2 (\partial_x h)(\partial_x
z)(\partial_y z) \right\}^{1/2}}.\label{hatT-full}
\end{equation}
We can also define the unit vector normal to the contact line (see
Fig.~\ref{fig:schematics}) as
\begin{equation}
{\hat {\bf n}}={1 \over \sqrt{1+(\partial_x h)^2}}(-\partial_x
h,1,0)),\label{n}
\end{equation}
using which we can define the local contact-angle as
\begin{equation}
\theta(x)=\cos^{-1} \left({\hat {\bf T}} \cdot {\hat {\bf
n}}\right),\label{theta-T-n}
\end{equation}
where
\begin{equation}
{\hat {\bf T}} \cdot {\hat {\bf n}}={\sqrt{1+(\partial_x h)^2}
\over \left\{1+(\partial_y z)^2+(\partial_x h )^2
\left[1+(\partial_x z)^2\right]-2 (\partial_x h)(\partial_x
z)(\partial_y z) \right\}^{1/2}}.\label{T-full}
\end{equation}
Note that we have $\sqrt{1+(\partial_x h)^2} {\hat {\bf T}} \cdot
{\hat {\bf y}}=\cos \theta(x)$.
\begin{multicols}{2}\noindent

\section{Coupling Constants in the Two Different Approaches}  \label{aCouplings}

In this Appendix, we have summarized the limiting form of the
coupling constants of the dynamical equation, corresponding to
each of the dissipation mechanisms separately.

\subsection{Local Approach}  \label{slocalDE}

The asymptotic form of the equations given in Sec. \ref{sEquation}
can be obtained in this case by taking the limit $c_{0h}/c_{0l}
\to \infty$. One obtains
\begin{equation}
\left.v\right|_l={c_{0l} \over 2}\left(1-{\theta^2 \over
\theta_e^2}\right), \label{v-theta-l}
\end{equation}
for the contact-angle--velocity relation,
\begin{equation}
c_l=b_0 c_{0l} {\theta^2 \over \theta_e^2}, \label{c-l}
\end{equation}
for the relaxation speed, and
\begin{equation}
\lambda_{1l}={c_{0l} \over 2}\left[(1-2 b_1){\theta^2 \over
\theta_e^2}-1\right], \label{lambda1-l}
\end{equation}
and
\begin{equation}
\lambda_{2l}=(b_2+b_0^2) c_{0l} {\theta^2 \over \theta_e^2},
\label{lambda2-l}
\end{equation}
and
\begin{equation}
\lambda_{3l}=b_3 c_{0l} {\theta^2 \over \theta_e^2},
\label{lambda3-l}
\end{equation}
for the nonlinear coupling constants. We can also find the
strength of the noise term as
\begin{equation}
D_{l}={a^2 c_{0l} \over \left|1-{\theta^2 \over
\theta_e^2}\right|} \left(g \over \gamma \theta_e^2\right)^2.
\label{D-l}
\end{equation}

\subsection{Hydrodynamic Approach}  \label{shydroDE}

The corresponding asymptotic forms of the equations given in Sec.
\ref{sEquation} in this case can be obtained by taking the limit
$c_{0h}/c_{0l} \to 0$. One obtains
\begin{equation}
\left.v\right|_h={c_{0h} \over 2}{\theta \over
\theta_e}\left(1-{\theta^2 \over \theta_e^2}\right),
\label{v-theta-h}
\end{equation}
for the contact-angle--velocity relation,
\begin{equation}
c_h=b_0 {c_{0h} \over 2}{\theta \over \theta_e} \left(3{\theta^2
\over \theta_e^2}-1\right), \label{c-h}
\end{equation}
for the relaxation speed, and
\begin{equation}
\lambda_{1h}={c_{0h} \over 2}{\theta \over
\theta_e}\left[(b_1-1)+(1-3 b_1){\theta^2 \over
\theta_e^2}\right], \label{lambda1-h}
\end{equation}
and
\begin{equation}
\lambda_{2h}={c_{0h} \over 2}{\theta \over \theta_e} \left[-b_2+(3
b_2+6 b_0^2){\theta^2 \over \theta_e^2}\right], \label{lambda2-h}
\end{equation}
and
\begin{equation}
\lambda_{3h}=b_3 {c_{0h} \over 2}{\theta \over \theta_e}
\left(3{\theta^2 \over \theta_e^2}-1\right), \label{lambda3-h}
\end{equation}
for the nonlinear coupling constants. We can also find the
strength of the noise term as
\begin{equation}
D_{h}={a^2 c_{0h} \over \left|1-{\theta^2 \over
\theta_e^2}\right|} \left({\theta \over \theta_e}\right) \left(g
\over \gamma \theta_e^2\right)^2. \label{D-h}
\end{equation}

\end{multicols}

\begin{references}
\bibitem{notation}
Note that in this section, we are dealing with $d$-dimensional
wavevectors, and we are using the notation $k=|{\bf k}|$ for
simplicity. This is to be contrasted from previous sections where
$k$ stands for a one dimensional vector taking both positive and
negative values.
\end{references}
\end{document}